\documentclass[aps,prd,showpacs,nofootinbib,twocolumn]{revtex4}
\usepackage{amsmath,amssymb}
\usepackage[dvips]{graphicx}
\usepackage{dcolumn}% Align table columns on decimal point
\usepackage{bm}% bold math

\newcommand{\nc}[1]{\newcommand{#1}}

\nc{\its}[1]{\itshape #1 \upshape}
\nc{\mc}[3]{\multicolumn{#1}{#2}{#3}}

\nc{\bc}{\begin{center}}
\nc{\ec}{\end{center}}
\nc{\ig}[1]{\bc \includegraphics{#1} \ec}

\nc{\ben}{\begin{enumerate}}
\nc{\een}{\end{enumerate}}

\nc{\bo}[1]{\mbox{\boldmath \( #1 \! \! \)  \unboldmath}}

\nc{\be}{\begin{eqnarray}}
\nc{\ee}{\end{eqnarray}}
\nc{\bew}{\begin{eqnarray*}}
\nc{\eew}{\end{eqnarray*}}

\nc{\nnn}{\nonumber}
\nc{\f}[2]{\frac{#1}{#2}}
\nc{\td}[2]{\f{d #1}{d #2}}
\nc{\pd}[2]{\f{\partial #1}{\partial #2}}
\nc{\suli}{\sum\limits}
\nc{\proli}{\prod\limits}
\nc{\ili}{\int\limits}
\nc{\sr}[2]{\stackrel{#1}{#2}}
\nc{\dps}{\displaystyle}
\nc{\ket}[1]{\left| #1 \right>}
\nc{\bra}[1]{\left< #1 \right|}
\nc{\bracket}[2]{\left< #1 \right| \left. \! #2 \right>}
\nc{\norm}[1]{\left\| #1 \right\|}
\nc{\lndm}[1]{\pd{^{#1} \ln{\det{D}}}{\mu^{#1}}}
\nc{\pdmm}[1]{D^{-1} \pd{^{#1} D}{\mu^{#1}}}
\nc{\pdm}{D^{-1}\pd{D}{\mu}}
\nc{\trac}[1]{\mbox{Tr}\left(#1\right)}

\nc{\la}{\langle}
\nc{\ra}{\rangle}
\def\simge{\mathrel{%
       \rlap{\raise 0.511ex \hbox{$>$}}{\lower 0.511ex \hbox{$\sim$}}}}
\def\simle{\mathrel{
       \rlap{\raise 0.511ex \hbox{$<$}}{\lower 0.511ex \hbox{$\sim$}}}}
\nc{\Vsi}{V_{\bf si}}
\nc{\Vav}{V_{\bf av}}
\nc{\tr}{{\rm Tr}}
\nc{\bx}{{\bf x}}
\nc{\by}{{\bf y}}
\nc{\bz}{{\bf 0}}
\nc{\Tpc}{T_{\rm pc}}
\nc{\mpsmv}{ m_{_{\rm PS}}/m_{_{\rm V}} }
\nc{\dg}{\dagger}
\nc{\Qb}{\bar{Q}}
\nc{\mq}{\mu_{q}}
\nc{\Te}{{\cal T}}
\nc{\Ca}{{\cal C}}
\nc{\C}{C}
\nc{\R}{{\cal R}}

\nc{\Om}{\Omega_{\rm M}}
\nc{\Oe}{\Omega_{\rm E}}
\nc{\Ome}{\Omega_{\rm M+}}
\nc{\Omo}{\Omega_{\rm M-}}
\nc{\Oee}{\Omega_{\rm E+}}
\nc{\Oeo}{\Omega_{\rm E-}}

\nc{\Cmef}{\tilde{C}_{{\rm M}+}}
\nc{\Cmof}{\tilde{C}_{{\rm M}-}}
\nc{\Ceef}{\tilde{C}_{{\rm E}+}}
\nc{\Ceof}{\tilde{C}_{{\rm E}-}}
\nc{\Cmei}{C_{{\rm M}+}}
\nc{\Ceoi}{C_{{\rm E}-}}

\nc{\Cmf}{\tilde{C}_{{\rm M}}}
\nc{\Cef}{\tilde{C}_{{\rm E}}}
\nc{\Cmi}{C_{{\rm M}}}
\nc{\Cei}{C_{{\rm E}}}

\nc{\Mmei}{m_{_{\rm M+}}}
\nc{\Meoi}{m_{_{\rm E-}}}
\nc{\Mmef}{\tilde{m}_{{\rm M}+}}
\nc{\Mmof}{\tilde{m}_{{\rm M}-}}
\nc{\Meef}{\tilde{m}_{{\rm E}+}}
\nc{\Meof}{\tilde{m}_{{\rm E}-}}

\nc{\Cav}{C_{\bf av}}
\nc{\Csi}{C_{\bf si}}
\begin{document}

%%%%%%%%%%%%%%%%%%%%%%%%%%%%%%%%%%%%%%%
%\baselineskip 16pt plus 1pt minus 1pt %
%%%%%%%%%%%%%%%%%%%%%%%%%%%%%%%%%%%%%%%

\preprint{\sf TKYNT-10-01, UTHEP-605, \ \ 2010/March}

\title{Electric and Magnetic Screening Masses at Finite Temperature\\
from Generalized Polyakov-Line Correlations in Two-flavor Lattice QCD}

\author{Y.~Maezawa$^1$,
S.~Aoki$^2$, S.~Ejiri$^3$, T.~Hatsuda$^4$, N.~Ishii$^4$,
 K.~Kanaya$^2$, N.~Ukita$^5$ and T.~Umeda$^6$ \\
(WHOT-QCD Collaboration)}

\affiliation{
$^1$En'yo Radiation Laboratory, Nishina Accelerator Research Center, RIKEN, Wako, Saitama 351-0198, Japan \\
$^2$Graduate School of Pure and Applied Sciences, University of Tsukuba, Tsukuba, Ibaraki 305-8571, Japan \\
$^3$Graduate School of Science and Technology, Niigata University, Niigata 950-2181, Japan \\
$^4$Department of Physics, The University of Tokyo, Tokyo 113-0033, Japan \\
$^5$Center for Computational Sciences, University of Tsukuba, Tsukuba, Ibaraki 305-8577, Japan \\
$^6$Graduate School of Education, Hiroshima University, Hiroshima 739-8524, Japan}

\date{\today}

\begin{abstract}
Screenings  of the quark-gluon plasma in electric and magnetic sectors 
are studied on the basis of generalized  Polyakov-line correlation functions
 in lattice QCD simulations with two flavors of improved Wilson quarks. 
Using the Euclidean-time reflection ($\R$) 
 and the charge conjugation ($\Ca$), electric and magnetic
   screening masses are extracted in a gauge 
  invariant manner. Long distance behavior of the standard Polyakov-line
  correlation in the quark-gluon plasma 
  is found to be dictated  by the magnetic screening.
 Also, ratio of the two screening masses 
 agrees with that obtained from the dimensionally-reduced effective field theory
   and the ${\cal N}=4$ supersymmetric Yang-Mills theory. 
\end{abstract}


\maketitle

%%%%%%%%%%%%%%%%%%%%%%%%%%%%%%%%%%%%%%%%%%%%%%%%%%%%%%%%%%%%%%%%%%%%%%
\section{Introduction}
\label{sec:intro}

Screenings inside the quark-gluon plasma (QGP)
 at high temperature are dictated by thermal fluctuations of 
  quarks and gluons,
  and are characterized by the electric and magnetic screening masses 
  \cite{Kraemmer:2003gd}. 
From the phenomenological point of view,
 screening of color charge affects the force between heavy-quarks 
  and determines the fate of heavy-quark bound states
    such as $J/\psi$ and $ \Upsilon$  
     inside hot QCD matter produced in  relativistic
     heavy-ion collisions at  
     RHIC and  LHC \cite{Matsui-Satz,QQ-rev}.

 Non-perturbative determinations of the screening masses
 have been attempted from various approaches in the past:
  Use of  Polyakov-line correlations with appropriate
  projection in color space 
  \cite{Nadkarni:1986cz,Kaczmarek:2004gv,Nakamura:2004wra,Maezawa:2007fc},
   use of  correlations of spatially local operators
  classified in terms of  Euclidean time-reflection \cite{Arnold:1995bh},
   use of  dimensionally-reduced effective field theory \cite{Hart:2000ha,
  Laine:2009dh}
  and a direct lattice evaluation of the gluon propagator under 
   gauge fixing \cite{Nakamura:2003pu}.
 The screening masses of the ${\cal N}=4$ supersymmetric Yang-Mills theory
 for large 't Hooft coupling  have been also 
 analysed on the basis of the AdS/CFT correspondence \cite{Bak:2007fk}.

In this paper, we extract
  electric and magnetic screening masses 
 from  gauge-invariant Polyakov-line correlations 
 in two-flavor lattice QCD simulations (a preliminary account has been
  reported in \cite{Maezawa:2008kh}). 
According to the symmetry properties under
  the Euclidean time-reflection $\R$  
  and the charge conjugation $\Ca$  \cite{Arnold:1995bh}, 
   we decompose the standard Polyakov-line operator $\Omega$ into
    four independent types, $\Omega_{{\rm M}\pm}$ and $\Omega_{{\rm E}\pm}$.
   Here M and E stand for the $\R$-even magnetic sector 
    and the $\R$-odd electric sector, respectively, while  $\pm$ stands for
     the even or odd under $\Ca$. Then the magnetic (electric) screening
      mass is extracted from the 
      correlation of Tr$\Omega_{{\rm M}+}$  (Tr$\Omega_{{\rm E}-}$). 
 In our simulations on a $16^3 \times 4$ lattice,
we employ  two-flavor lattice QCD with improved Wilson quarks coupled
 to an RG-improved glue (Iwasaki gauge action) for
 the temperatures $T/\Tpc \simeq 1$--4 with 
 $\Tpc$ being the pseudocritical temperature.  
 We take the quark masses corresponding to 
 $\mpsmv = 0.65$ and 0.80 
  where $m_{\rm PS}$ ($m_{\rm V}$) is pseudoscalar (vector) meson mass at $T=0$. 
 
This paper is organized as follows:
In Sec.~\ref{sec:2}, we discuss the properties 
 of the Polyakov-line operators under $\R$ and $\Ca$.
Numerical simulations and the results of
 the screening masses are shown in Sec.~\ref{sec:3}.
 Comparison of  our results with those
  obtained from 
  the dimensionally-reduced effective field theory and 
   the ${\cal N}=4$ supersymmetric Yang-Mills theory are also given.
   Section \ref{sec:summary} is devoted to summary and concluding remarks.

%%%%%%%%%%%%%%%%%%%%%%%%%%%%%%%%%%%%%%%%%%%%%%%%%%%%%%%%%%%%%%%%%%%%%%%%%%

\section{Generalized Polyakov-line correlations}
\label{sec:2}

We start with the standard Polyakov-line operator 
\be
\Omega (\bx) = P \exp \left[ i g \int_0^{1/T} d\tau A_4(\tau,\bx) \right] .
\ee
Its gauge invariant correlation function reads
\be
C_{\Omega} (r,T) \equiv 
\langle \tr \Omega^\dagger ({\bf x}) \tr \Omega ({\bf y}) \rangle 
- |\langle \tr\Omega \rangle|^2,
\label{eq:Cav} 
\ee
 where $r \equiv |\bx - \by|$ and we define 
$\tr \Omega \equiv \f{1}{N_c} \sum_{\alpha=1}^{N_c} \Omega^{\alpha \alpha}$ with $\alpha$ being color indices.
The  screening mass is defined from the exponential fall-off
 of this gauge invariant correlation,
 \be
C_{\Omega}(r,T) \xrightarrow[r\rightarrow \infty]{ } 
\gamma_{_{\Omega}}(T) \frac{e^{- m_{_{\Omega}}(T) r}}{rT}.
\label{eq:sPm}
\ee
Since the magnetic gluon is less screened at finite $T$,
 the dominant contribution to $m_{_{\Omega}}(T)$ would come from
 non-perturbative magnetic sector of QCD as pointed out
 in \cite{Arnold:1995bh}.

In order to extract the electric and magnetic screening masses separately,
 we  classify the Polyakov-line operator into different classes 
 on the basis of the symmetries under
  $\R$ and $\Ca$ which 
 are good symmetries of QCD at zero chemical potential \cite{Arnold:1995bh}.
 Under $\R$ and $\Ca$, the gluon fields transform as,
\be
 {A}_i(\tau,\bx) &\xrightarrow{\R}& {A}_i(-\tau,\bx) ,\ \
 A_4 (\tau,\bx)    \xrightarrow{\R}   - A_4(-\tau,\bx) ,\\
 A_\mu(\tau,\bx) &\xrightarrow{\Ca}& - A^*_\mu(\tau,\bx) .
\label{eq:TeA}
\ee
We call an operator  magnetic (electric) 
if it is even (odd) under $\R$. 
 Under these transformations, the standard Polyakov-line operator
transforms as
\be
\Omega \xrightarrow{\R} \Omega^\dag,
\quad
\Omega \xrightarrow{\Ca} \Omega^\ast
.
\ee
Then, we can define magnetic (electric) operator which is
$\R$-even ($\R$-odd) as
\be
\Om \equiv \f{1}{2} (\Omega + \Omega^\dagger),
\quad
\Oe \equiv \f{1}{2} (\Omega - \Omega^\dagger).
\ee
 Furthermore, they can be decomposed into  $\Ca$-even and $\Ca$-odd
  operators as
\be
\Omega_{{\rm M} \pm} = \frac{1}{2} (\Om \pm \Om^* )
\quad
\Omega_{{\rm E} \pm} = \frac{1}{2} (\Oe \pm \Oe^* )
 . \label{eq:Omega-E} 
\ee

Using these operators and noting the fact that 
  $\tr \Omo = \tr \Oee = 0$, we can define
   two generalized gauge-invariant Polyakov-line correlation functions,
\be
C_{\rm M +} (r,T) &\equiv&
 \langle \tr \Ome ({\bf x}) \tr \Ome ({\bf y}) \rangle 
 - |\langle \tr\Omega \rangle|^2
, \label{eq:Cmp} \\
C_{\rm E-} (r,T) &\equiv&
 \langle \tr \Oeo ({\bf x}) \tr \Oeo ({\bf y}) \rangle
. \label{eq:Cmm}
\ee
 Note that $\tr \Ome$ 
 ($\tr \Oeo $) is nothing but the real (imaginary) part
 of  $\tr \Omega$, and that $\langle \tr \Omega \rangle$ is
real due to the $\Ca$ symmetry.

 The electric and magnetic screening masses
 can be defined from the above gauge invariant correlation functions
 through their long distance behavior:
\be
C_{\rm M+} (r,T)&  \xrightarrow[r \rightarrow \infty]{ }&
 \gamma_{_{\rm M+}}(T) \f{e^{-m_{\rm M+}(T) r} }{rT} ,
\label{eq:SCF0} \\
 C_{\rm E-} (r,T)&  \xrightarrow[r \rightarrow \infty]{ }&
 \gamma_{_{\rm E-}}(T) \f{e^{-m_{\rm E-}(T) r} }{rT},
\label{eq:SCF}
\ee 

Notice that the standard Polyakov-line correlation function and the 
above generalized correlation functions are simply related as
\be
C_{\Omega}(r,T)  = C_{\rm M+} (r,T)  - C_{\rm E-} (r,T) .
\ee
Thus the separate determination of 
 $C_{\rm M+} (r,T)$ and $C_{\rm E-} (r,T)$ on the lattice 
 enables us to study the 
  relative importance of the magnetic and electric sector in 
 a non-perturbative manner. 

%%%%%%%%%%%%%%%%%%%%%%%%%%%%%%%%%%%%%%%%%%%%%%%%%%%%%%%%%%%%%%%%%%%%%

\section{Lattice simulations}
\label{sec:3}

\subsection{Lattice setup}

We employ a renormalization group improved gauge action $S_g$
 and a clover improved
Wilson quark action with two flavors  $S_q$:
\begin{eqnarray}
  S_g =
  -{\beta}\sum_x \Big(
   c_0 \!\!\!\!\!\!\!\!  \sum_{\mu<\nu;\mu,\nu=1}^{4} \!\!\!\!\!\! W_{\mu\nu}^{1\times1}(x) 
  +c_1 \!\!\!\!\!\!\!\! \sum_{\mu\ne\nu;\mu,\nu=1}^{4} \!\!\!\!\!\! W_{\mu\nu}^{1\times2}(x) \Big), \\
  S_q = \sum_{f=1,2}\sum_{x,y}\bar{q}_x^f D_{x,y}q_y^f,
\end{eqnarray}
where 
$\beta=6/g^2$, $c_1=-0.331$, $c_0=1-8c_1$, and  $W_{\mu\nu}^{n \times m}$
 is a $n \times m$ rectangular shaped Wilson loop and
\begin{eqnarray}
 D_{x,y} = \delta_{xy}
   -{K}\sum_{\mu} \{ (1-\gamma_{\mu})U_{x,\mu}\delta_{x+\hat{\mu},y}
  \nnn \\ 
   +(1+\gamma_{\mu})U_{x,\mu}^{\dagger}\delta_{x,y+\hat{\mu}}\}
   -\delta_{xy}{c_{SW}}{K}\sum_{\mu<\nu}\sigma_{\mu\nu}F_{\mu\nu}.
\end{eqnarray}
Here $K$ is the hopping parameter and 
$F_{\mu\nu}$ is the lattice field strength,
$F_{\mu\nu} = {1}/{8i}(f_{\mu\nu}-f^{\dagger}_{\mu\nu})$,
 with $f_{\mu\nu}$ being the standard clover-shaped combination of gauge links.
For the clover coefficient $c_{SW}$, we adopt the mean field value using
$W^{1\times 1}$ which was calculated in the one-loop perturbation theory, 
\begin{eqnarray}
 {c_{SW}}=(W^{1\times 1})^{-3/4}=(1-0.8412\beta^{-1})^{-3/4}.
\end{eqnarray}
Our simulations are performed on a lattice with a size of 
$N_s^3 \times N_t = 16^3 \times 4$
 along lines of constant physics, 
i.e. lines of constant $\mpsmv$
(the ratio of pseudoscalar and vector meson masses) at $T=0$
in the space of simulation parameters.
Details on the lines of constant physics and the phase diagram for $N_t=4$
with the same actions as above are
given  in Refs.~\cite{cp1,Maezawa:2007fc}.
We take two values, $\mpsmv= 0.65$ and 0.80,
with the temperature range of $T/ \Tpc \sim$ 1.0--4.0 (10 points)
and 1.0--3.0 (7 points), 
 respectively, where $\Tpc$ is the pseudocritical 
temperature along the line of constant physics \cite{cp1,Maezawa:2007fc}.
The number of trajectories for each run after thermalization is 
5000--6000, and we measure physical quantities at every 10 trajectories.
The statistical errors are estimated by a jackknife method with
 a bin size of 100 trajectories.

\subsection{Screening masses}

\begin{figure}[tbp]
  \begin{center}
    \includegraphics[width=80mm]{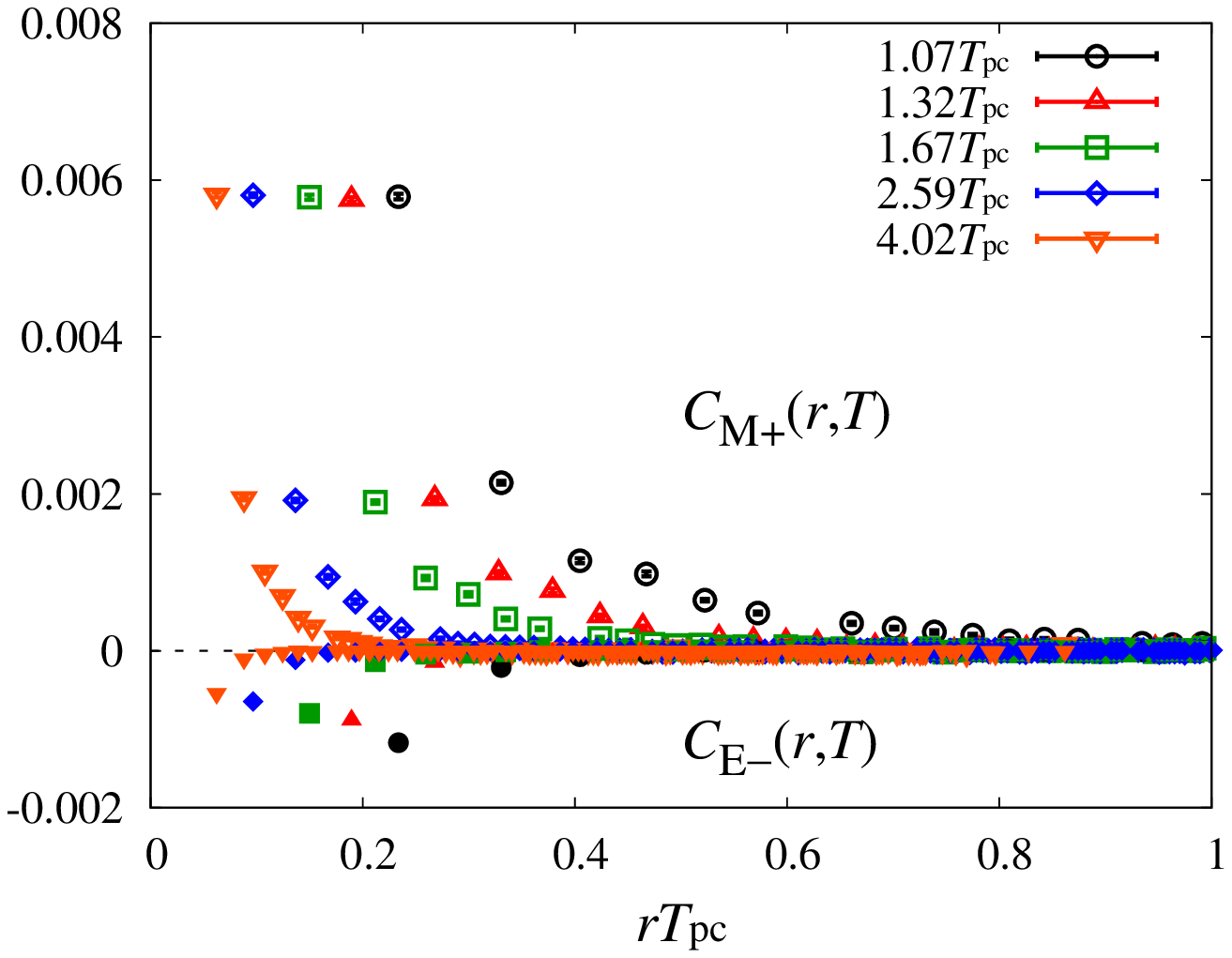} 
    \includegraphics[width=80mm]{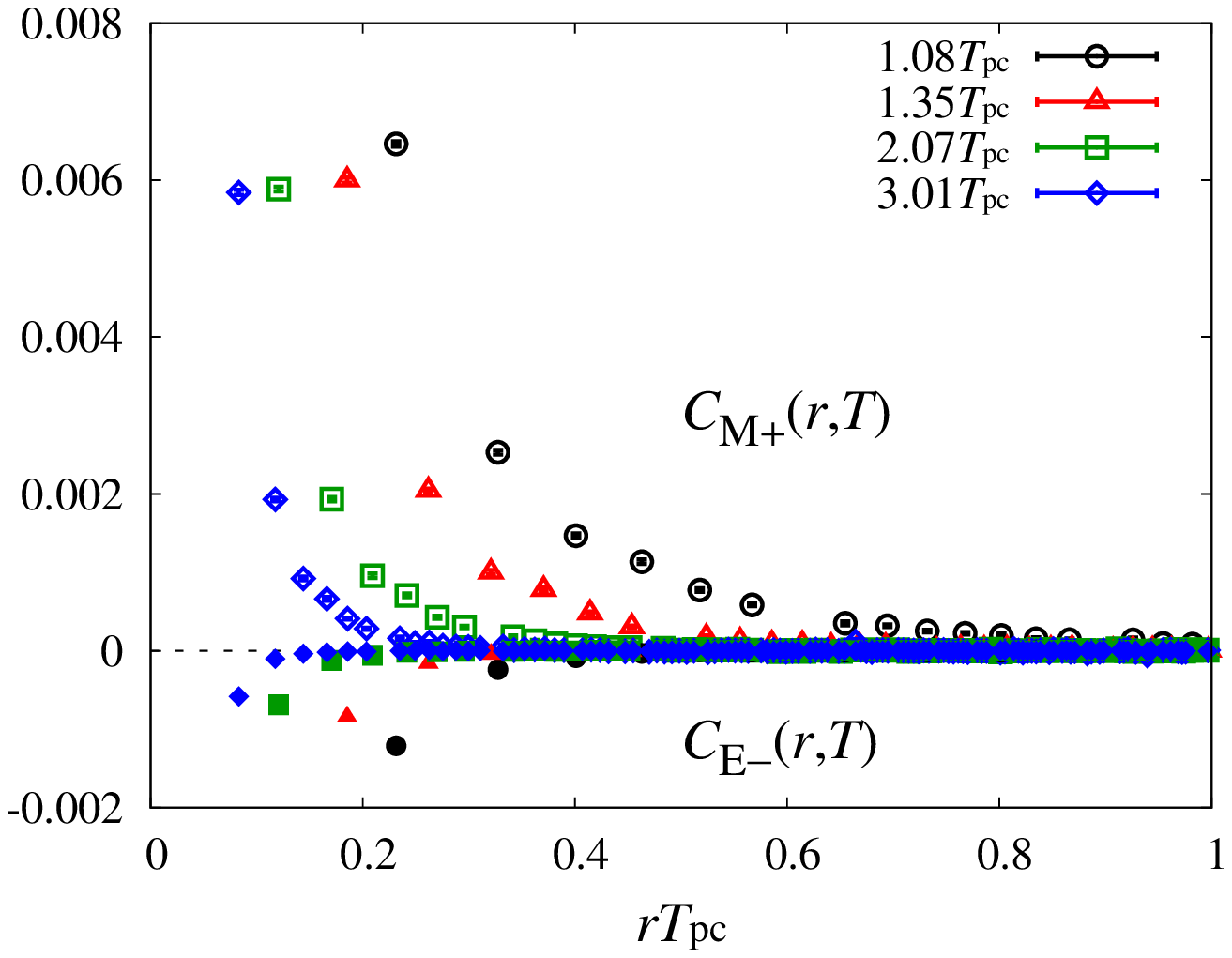}
    \caption{Results of $C_{\rm M+}(r,T)$ and $C_{\rm E-}(r,T)$
    for several temperatures as a function of $r \Tpc$
    at $\mpsmv = 0.65$ (upper panel) and 0.80 (lower panel).}
    \label{fig:CI}
  \end{center}
\end{figure}

\begin{figure}[tbp]
  \begin{center}
    \includegraphics[width=80mm]{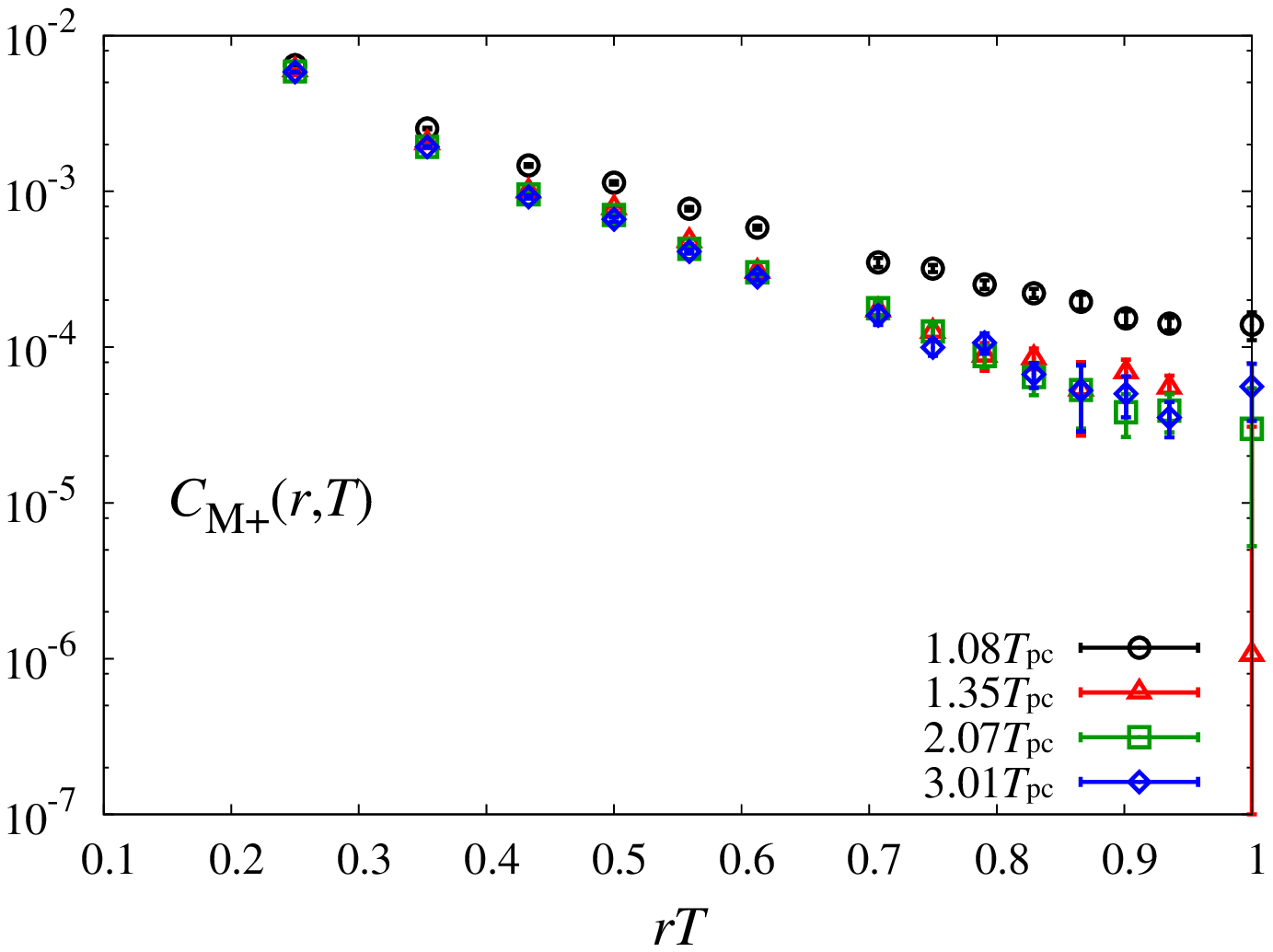} 
    \includegraphics[width=80mm]{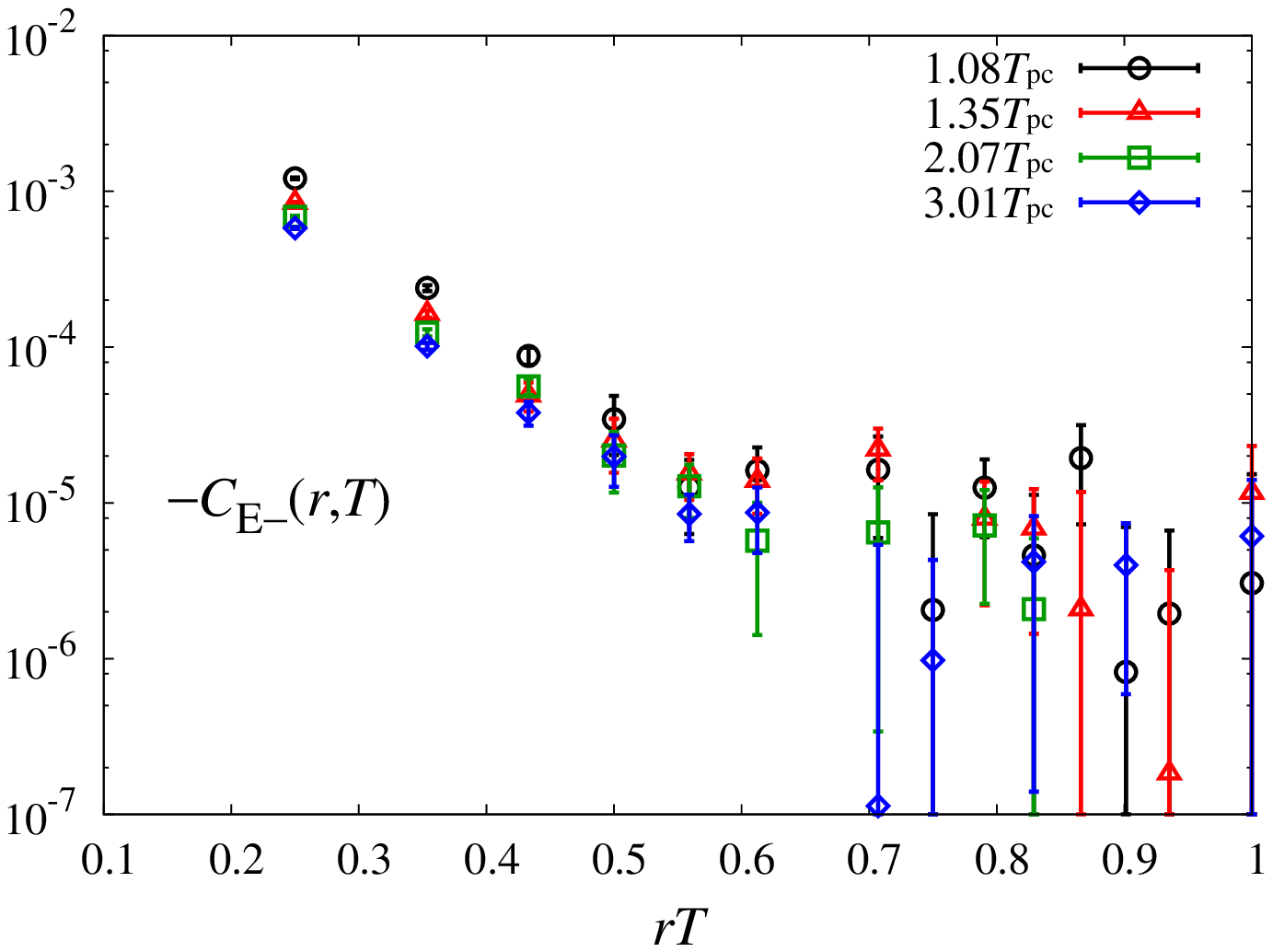}
    \caption{Results of $C_{\rm M+}(r,T)$ (upper panel) 
    and $-C_{\rm E-}(r,T)$ (lower panel) in a logarithmic scale 
     as a function of $r T$
    at $\mpsmv = 0.80$.}
    \label{fig:CI_log}
  \end{center}
\end{figure}

Results of the generalized Polyakov-line correlations $C_{\rm M+(E-)}(r,T)$
 are shown in Fig.~\ref{fig:CI}
  for  $\mpsmv =0.65$ (upper panel) and 0.80 (lower panel).
These figures show that (i) the magnetic 
correlation and the electric correlation have
 an opposite sign, and (ii) the magnetic correlation
 has larger magnitude and longer range than  the electric correlation at long
   distances. The latter implies that the 
   standard Polyakov-line correlation is dominated 
    by the magnetic sector.
In Fig.~\ref{fig:CI_log} we show $C_{\rm M+}(r,T)$ (upper panel)
 and $-C_{\rm E-}(r,T)$ (lower panel) in a logarithmic scale as a function of
  $rT$ for $\mpsmv=0.80$. 
  These correlators behave linearly and 
   are scaled well by $rT$ at high temperature, so that
 the screening masses can be extracted 
 by fitting the correlators using the Yukawa form 
 defined in Eqs.~\eqref{eq:SCF0} and \eqref{eq:SCF}.

\begin{figure}[tb]
  \begin{center}
    \includegraphics[width=80mm]{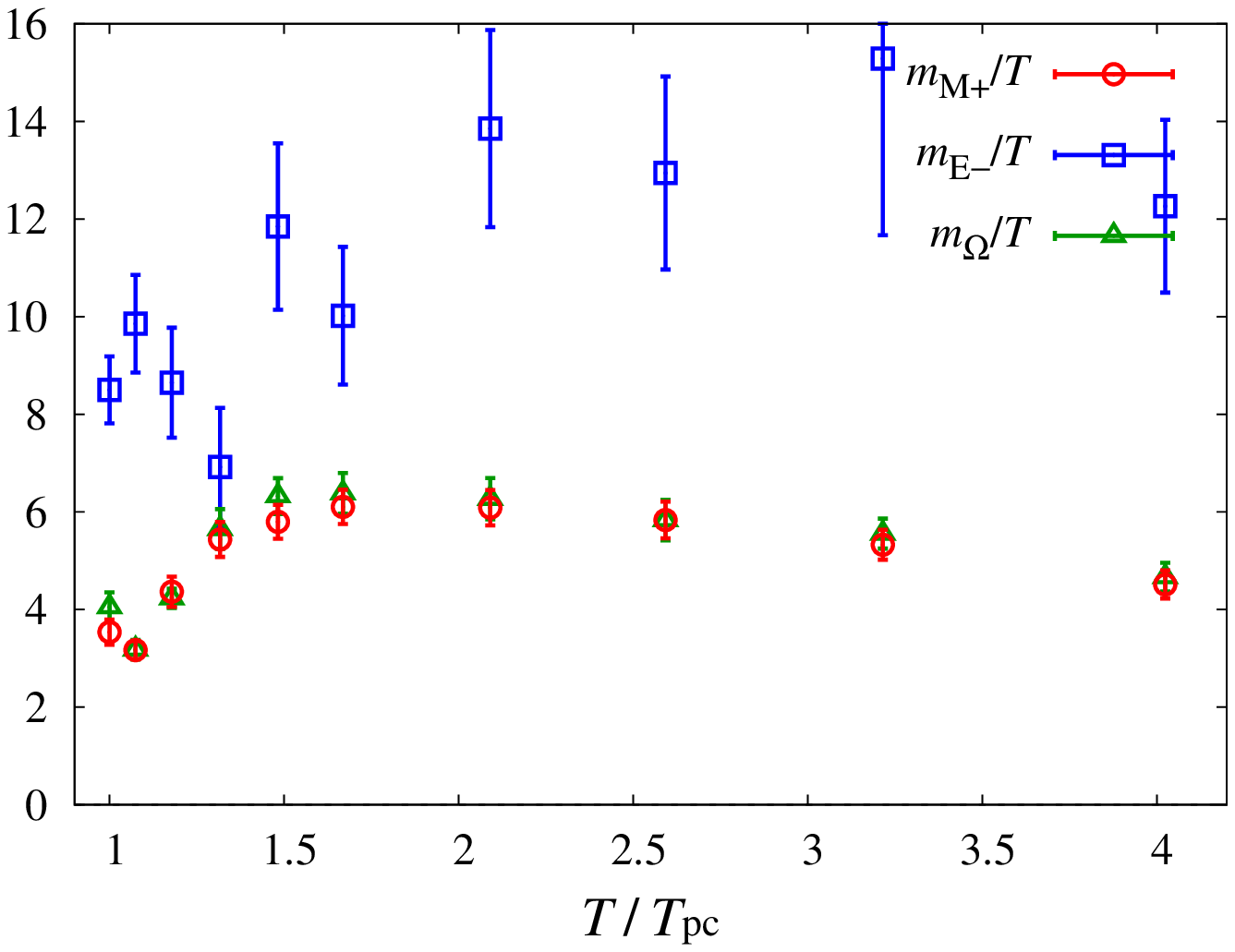} 
    \includegraphics[width=80mm]{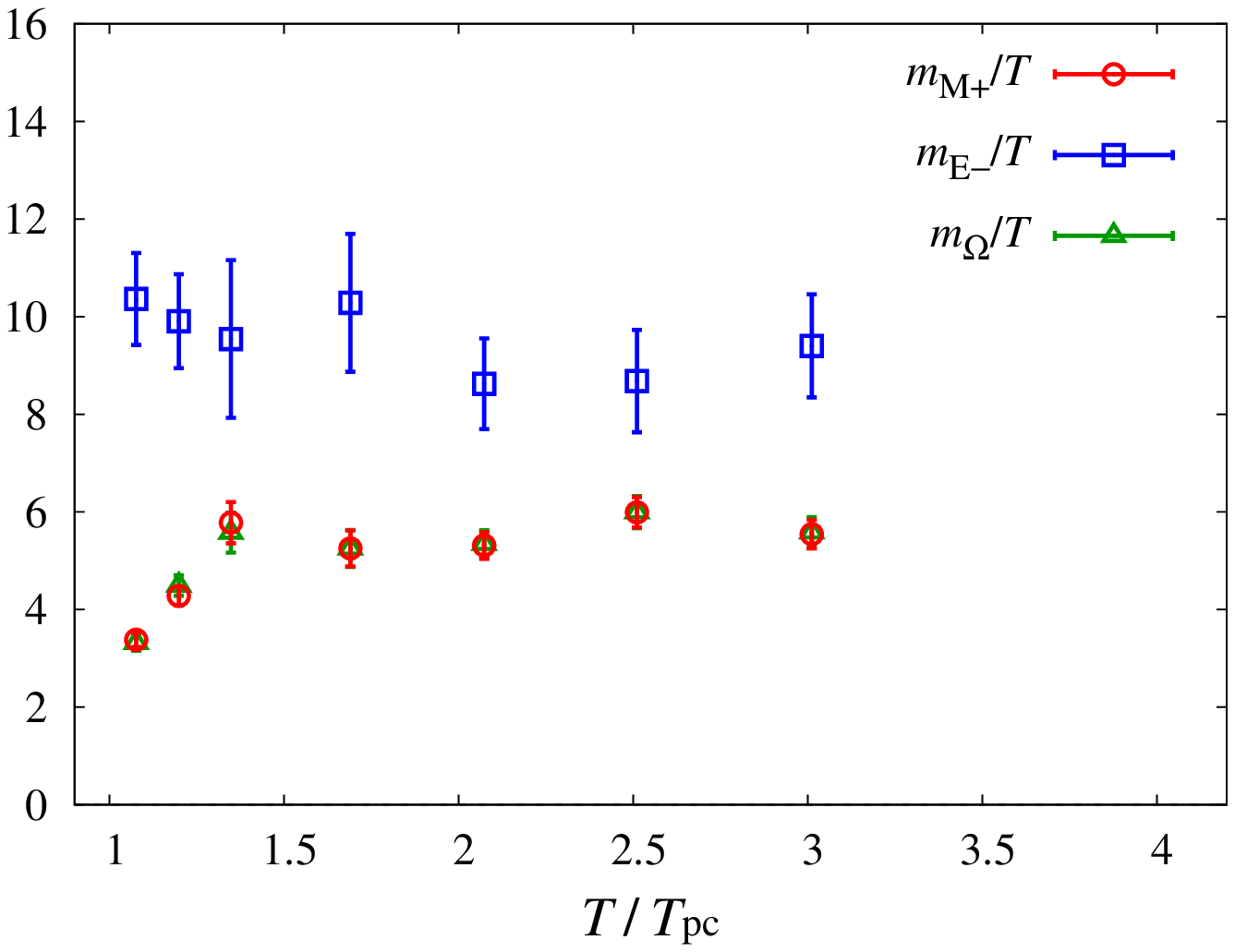} 
    \caption{Results of magnetic and electric screening mass
    $\Mmei/T$ and $\Meoi/T$ together with the screening mass obtained from the 
    standard Polyakov-line correlation
    $m_{_{\Omega}}/T$  as a function of temperature
    at $\mpsmv = 0.65$
    (upper panel) and  0.80 (lower panel). }
    \label{fig:SM}
  \end{center}
\end{figure}

 We fit these correlators by Eqs.~\eqref{eq:SCF0} and \eqref{eq:SCF}
 in the same interval of $0.5 \le rT \le 1.0$
  by minimizing $\chi^2/N_{\rm dof}$.\footnote{
 We study the fit range dependence of the results, and find that
 the magnitude of systematic error due to the fit range is smaller than or
  comparable to the statistical errors at $T \simge 1.2 \Tpc$.}
 Upper (lower) panel of Fig.~\ref{fig:SM} shows results of screening masses
 in the magnetic and electric sector at $\mpsmv = 0.65$ (0.80)
 as a function of temperature.
Numerical results are summarized in Tab.~\ref{tab:SM_065} and \ref{tab:SM_080}
 for $\mpsmv = 0.65$ and 0.80, respectively. The screening mass obtained from the 
 standard Polyakov-line correlation function $m_{_\Omega}(T)$ using Eq.(\ref{eq:sPm}) 
 is also shown.
 As expected, the magnetic sector has longer range than the electric sector 
 ($m_{\rm M+}(T)$ is smaller than $m_{\rm E-}(T)$) and also
  the standard screening mass is dominated by the magnetic mass
  ($m_{_\Omega}(T) \simeq m_{\rm M+}(T)$).

\begin{table}[h]
 \begin{center}
 \caption{Results of screening masses at $\mpsmv = 0.65$.
 First parentheses show statistical errors, while
 the second parentheses show systematic errors
  calculated from difference between screening masses
  in the range of $0.5 < rT < 1.0$ and $\sqrt{6}/4 < rT < 1.0$.}
 \label{tab:SM_065}
 {\renewcommand{\arraystretch}{1.2} \tabcolsep = 3mm
 \newcolumntype{.}{D{.}{.}{6}}
 \begin{tabular}{c|c|c}
 \hline\hline
 \multicolumn{1}{c|} {$T/T_{\rm pc}$} &
 \multicolumn{1}{c|} {$\Mmei/T$} &    
 \multicolumn{1}{c} {$\Meoi/T$}  \\ 
 \hline
 1.00 &  3.5( 2)( 1) &    8.5( 6)(17)  \\
 1.07 &  3.2( 1)( 3) &    9.9(10)( 0)  \\
 1.18 &  4.4( 3)( 1) &    8.6(11)( 0)  \\
 1.32 &  5.4( 3)( 4) &    6.9(12)(23)  \\
 1.48 &  5.8( 3)( 2) &   11.8(17)(14)  \\
 1.67 &  6.1( 3)(13) &   10.0(14)( 7)  \\
 2.09 &  6.1( 3)( 5) &   13.9(20)( 1)  \\
 2.59 &  5.8( 3)( 0) &   12.9(19)(45)  \\
 3.22 &  5.3( 3)( 2) &   15.3(36)(25)  \\
 4.02 &  4.5( 2)( 2) &   12.3(17)( 9)  \\
 \hline\hline
 \end{tabular}}
 \end{center}
\end{table}

\begin{table}[h]
 \begin{center}
 \caption{Results of screening masses at $\mpsmv = 0.80$
  with the errors calculated by the same procedure as Tab.~\ref{tab:SM_065}.
 }
 \label{tab:SM_080}
 {\renewcommand{\arraystretch}{1.2} \tabcolsep = 3mm
 \newcolumntype{.}{D{.}{.}{6}}
 \begin{tabular}{c|c|c}
 \hline\hline
 \multicolumn{1}{c|} {$T/T_{\rm pc}$} &
 \multicolumn{1}{c|} {$\Mmei/T$} &
 \multicolumn{1}{c} {$\Meoi/T$}  \\ 
 \hline
 1.08 &  3.4( 1)( 0) &    0.4( 9)( 0)  \\
 1.20 &  4.3( 1)( 1) &    9.9( 9)( 0)  \\
 1.35 &  5.8( 4)( 3) &    9.5(16)(42)  \\
 1.69 &  5.3( 3)( 2) &    0.3(14)( 2)  \\
 2.07 &  5.3( 2)( 1) &    8.6( 9)( 3)  \\
 2.51 &  6.0( 3)( 1) &    8.7(10)(10)  \\
 3.01 &  5.5( 2)( 0) &    9.4(10)( 5)  \\
 \hline\hline
 \end{tabular}}
 \end{center}
\end{table}

%%%%%%%%%%%%%%%%%%%%%%%%%%%%%%%%%%%%%%%%%%%%%%%%%%%%%%%%%
\subsection{Comparison to other approaches}

Let us compare our results with the predictions by
 the dimensionally-reduced effective field theory (3D-EFT) 
  and the ${\cal N}=4$ supersymmetric Yang-Mills theory (SYM).
 
 In the 3D-EFT approach \cite{Hart:2000ha}, the screening masses 
 in various channels with the quantum numbers $J_{\cal R}^{{\cal P} {\cal C}}$ are
  calculated: Here  ${\cal P}$ is a parity
  in a two-dimensional plane perpendicular to $\bx$-$\by$, and
  $J$ is an angular momentum in the two-dimensional plane. 
  Since the lowest masses for $(\R ,\Ca)=(+,+)$ and  $(\R ,\Ca)=(-,-)$ 
  are in $(J,{\cal P})=(0,+)$ channels, we extract
   $m_{\rm M+}$ from the $0_{+}^{++}$ channel and
   $m_{\rm E-}$  from the $0_{-}^{+-}$ channel.  
   Tables 6 and 8 of Ref.~\cite{Hart:2000ha} for
   the lattice coupling  $\beta=21$, lattice size $L=30$, $N_f=2$ and 
 $T\simeq 2\Lambda_{\overline{\rm MS}}$ lead to 
\be
& & \! \! \! \! \! \! \! \! \! \! \! \! \! \! \! \!
 \textrm{3D-EFT}\ (N_f=2):  \\
& & \ \ {\Mmei}/{T} = 3.96 (5) , \  {\Meoi}/{T} = 7.01 (10), \nonumber \\
& & \ \ {\Meoi}/{\Mmei} \simeq 1.77.\nonumber
\ee

The screening masses of the Polyakov-line correlation functions
  in ${\cal N}=4$ supersymmetric Yang-Mills theory 
 in the limit of large $N_c$ and 
   large 't Hooft coupling ($\lambda = g^2 N_c$) were calculated
    by  AdS/CFT correspondence \cite{Bak:2007fk}. Under  the same
   identification in $(J,{\cal P})=(0,+)$ channel as discussed above,
   Tab.~1 of Ref.~\cite{Bak:2007fk} leads to 
 \be
& & \! \! \! \! \! \! \! \! \! \! \! \! \! \! \! \!
 {\cal N}=4 \ \textrm{SYM}: \\
& & \ \ {\Mmei}/{T} = 7.34 , \  {\Meoi}/{T} = 16.05 ,\nonumber \\
& & \ \ {\Meoi}/{\Mmei} = 2.19. \nonumber
\ee

Results of our screening masses  
and their ratio for $1.67  < T/\Tpc < 3.22 $ with 
$\mpsmv=0.65$  in
   Tab.~\ref{tab:SM_065} indicate that
\be
& & \! \! \! \! \! \! \! \! \! \! \! \! \! \! \! \!
 \textrm{4D-lattice QCD} \ (N_f=2): \\
& & \ \ {\Mmei}/{T} = 5.8 (2) ,  \ {\Meoi}/{T} = 13.0 (11) , \nonumber\\
& & \ \ {\Meoi}/{\Mmei} = 2.3 (3).\nonumber
\ee

In all three cases above, we have an inequality $\Mmei < \Meoi$ so that
 the magnetic sector dominates at large distances.
  Since we cannot 
  compare the absolute magnitude
   of the screening masses in QCD and that in
  ${\cal N}=4$ SYM due to  different number of degrees of freedom,
 we take a ratio, $\Meoi/\Mmei$ (Table  \ref{tab:SMratio}), 
 and make a comparison    for the three cases in Fig.~\ref{fig:SMratio}.
 We find that  the ratio agrees well with each other
 for the temperature range of $1.5 < T/\Tpc < 3$.

\begin{table}[t]
\begin{center}
 \caption{Results of the screening ratio $\Meoi/\Mmei$
  in $N_f=2$ lattice QCD simulations.}
 \label{tab:SMratio}
{\renewcommand{\arraystretch}{1.2} \tabcolsep = 1.0mm
 \newcolumntype{.}{D{.}{.}{6}}
 \begin{tabular}{..|..}
\hline \hline
 \multicolumn{2}{c|}{$\mpsmv = 0.65$} & 
 \multicolumn{2}{c} {0.80} \\
 \hline
 \multicolumn{1}{c} {$T/\Tpc$} & \multicolumn{1}{c|} {$\Meoi/\Mmei$} & 
 \multicolumn{1}{c} {$T/\Tpc$} & \multicolumn{1}{c}  {$\Meoi/\Mmei$} \\
 \hline
 1.00 &  2.4( 2) & 1.08 &  3.1( 2)  \\
 1.07 &  3.1( 3) & 1.20 &  2.3( 2)  \\
 1.18 &  2.0( 3) & 1.35 &  1.7( 3)  \\
 1.32 &  1.3( 2) & 1.69 &  2.0( 3)  \\
 1.48 &  2.0( 3) & 2.07 &  1.6( 1)  \\
 1.67 &  1.6( 2) & 2.51 &  1.4( 1)  \\
 2.09 &  2.3( 3) & 3.01 &  1.7( 2)  \\
 2.59 &  2.2( 3) & &  \\
 3.22 &  2.9( 6) & &  \\
 4.02 &  2.7( 4) & &  \\
 \hline
\end{tabular}}
\end{center}
\end{table}

\begin{figure}[t]
  \begin{center}
    \includegraphics[width=80mm]{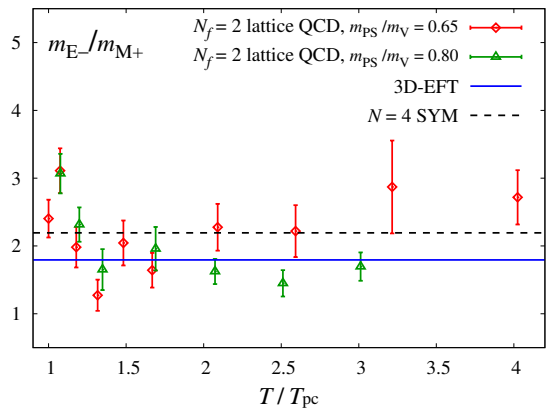} 
    \caption{Comparison of the screening ratio, $\Meoi/\Mmei$,
    with predictions in the dimensionally-reduced effective field theory (3D-EFT) \cite{Hart:2000ha}
    and ${\cal N}=4$ supersymmetric Yang-Mills theory (SYM) \cite{Bak:2007fk}. }
    \label{fig:SMratio}
  \end{center}
  \vspace{-5mm}
\end{figure}

%%%%%%%%%%%%%%%%%%%%%%%%%%%%%%%%%%%%%%%%%%%%%%%%%%%%%%%%%%%%%%%%%%%%%%
 
\ 

\ 

%%%%%%%%%%%%%%%%%%%%%%%%%%%%%%%%%%%%%%%%%%%%%%%%%%%%%%%%%%%%%%%%%%%%%%
\section{Summary}
\label{sec:summary}

We investigated the screening properties of the quark-gluon plasma
 from the Polyakov-line correlation functions classified under
 the Euclidean-time reflection ($\R$) symmetry 
  and the charge conjugation ($\Ca$) symmetry.
The Polyakov-line correlators with the quantum numbers 
 of $(\R,\Ca)=({\rm even},+)$ and $({\rm odd},-)$ are
 defined in gauge invariant form.

From the lattice simulations of $N_f=2$ QCD above $\Tpc$
 with the RG-improved gluon action and
  the clover-improved Wilson quark action,
   we extracted the magnetic screening mass $\Mmei$
   and the electric screening mass $\Meoi$, and found that
   ${\Meoi}/{\Mmei} = 2.1 (2)$ for $1 < \Tpc < 3$.
  Therefore, the standard Polyakov-line correlation functions at
   long distance is dictated by the magnetic screening.
 Our electric-magnetic ratio is consistent with those
  obtained from the dimensionally-reduced effective field theory (3D-EFT)
  and  the ${\cal N}=4$ supersymmetric Yang-Mills theory (SYM).
 
 In the present work, we have not attempted to make
 projections of the Polyakov-line operator $\Omega$ to the 
 operators with definite $J$ (two-dimensional angular momentum)
 and ${\cal P}$ (two-dimensional parity).  It would be a
  future task to make such projections, so that we can compare
 the 4D-lattice results with 3D-EFT and SYM in more details.
 In Ref.~\cite{Maezawa:2008kh}, we extracted
  four screening masses $\tilde{m}_{_{\rm X}}$  (${\rm X=M+,M-,E+,E-}$)
  from gauge-fixed correlation functions $\tilde{C}_{\rm X}(r,T)
  =\langle \tr [\Omega_{\rm X} ({\bf x})  \Omega_{\rm X} ({\bf y})] \rangle $;
  we found that the long distance behavior of the 
 color-singlet Polyakov-line correlation functions
    is dictated  by $\tilde{m}_{_{\rm E \pm}}$ which is smaller than
   $\Mmei$ and $\Meoi$ for $T > 1.5 \Tpc $.  Further studies
   are needed to have physical interpretation of this result.

%%%%%%%%%%%%%%%%%%%%%%%%%%%%%%%%%%%%%%%%%%%%%%%%%%%%%%%%%%%%%%%%%%%%%%
\section*{Acknowledgements}
We would like to thank M.~Laine for giving useful advice
 about the Euclidean time symmetry.
This work is in part supported 
by Grants-in-Aid of the Japanese Ministry
of Education, Culture, Sports, Science and Technology, 
(Nos.~17340066, 18540253, 19549001, 20340047, 21340049)
and by the Grant-in-Aid for Scientific
Research on Innovative Areas (No. 2004: 20105001, 20105003).
This work is in part supported 
also by the Large-Scale Numerical Simulation
Projects of CCS/ACCC, Univ.~of Tsukuba, 
and by the Large Scale Simulation Program of High Energy
Accelerator Research Organization (KEK) 
Nos.06-19, 07-18, 08-10 and 09-18.

%%%%%%%%%%%%%%%%%%%%%%%%%%%%%%%%%%%%%%%%%%%%%%%%%%%%%%%%%%%%%%%%%%%%%%

%%%%%%%%%%%%%%%%%%%%%%%%%%%%%%%%%%%%%%%%%%%%%%%%%%%%%%%%%%%%%%%%%%%%%%

\end{document}